\documentclass[hyper,notoc] {JHEP} 

\usepackage{epsfig}

\newcommand\fverb{\setbox\pippobox=\hbox\bgroup\verb}
\newcommand\fverbdo{\egroup\medskip\noindent%
			\fbox{\unhbox\pippobox}\ }
\newcommand\fverbit{\egroup\item[\fbox{\unhbox\pippobox}]}
\newbox\pippobox

\title{D-branes from $N$ non-BPS D9-branes in IIA theory }

\author{by J. Kluso\v{n}\\
	 Department of Theoretical Physics and Astrophysics\\
                   Faculty of Science, Masaryk University\\
Kotl\'{a}\v{r}sk\'{a} 2, 611 37, Brno\\
Czech Republic\\
	E-mail: \email{klu@physics.muni.cz}}

\preprint{\hepth{9910241}}	

\abstract{We construct unstable system $ D8+\overline{D8} $ as a kink
solution on world-volume of non-BPS D9-brane in IIA theory. Further we will make other checks confirming validity of our approach.  }


\def\tr{\mathrm{Tr}}

\begin{document}

\maketitle
\section{Introduction}

In the previous paper \cite{Kluson}, we have obtained   BPS Dp-brane in IIA theory from non-BPS D(p+1)-brane  via tachyon condensation in form of kink solution. In this article, we will continue in this approach   and we will see, that due to the tachyon condensation on $N$ non-BPS D9-branes in IIA theory, we can construct unstable system consisting D8-branes and D8-antibranes, confirming results in \cite{Horava}. The lower dimensional branes can then arise as 
tachyon kink solution on this system and in this ``step by step'' construction
we will be able to construct all D-branes in IIA theory. It is clear, that with
action for system brane and antibranes, that will be presented in this paper, 
we could construct all D-branes in IIB theory as well, following \cite{witen}.

The starting point in our analysis is action for non-BPS Dp-brane, proposed in 
ref.\cite{Sen}. Sen argued, that non-BPS D-brane must contain on its world-volume 32 fermionic degrees of freedom, because this brane completely breaks bulk s
persymmetry of IIA theory. The supersymmetric invariant action for BPS brane contains 32 fermionic degrees freedom, but half of them is unphysical due to the presence of local gauge symmetry on world-volume of BPS D-brane, $\kappa$ symmetry \cite{Aganagic}.We can expect, that non-BPS D-brane has the same DBI action as BPS brane, simply for the reason, that these branes have common origin and have the same spectrum of massless bosonic degrees of freedom on their world-volume.  DBI  term is manifestly invariant under space-time supersymmetry, which
should be hold for non-BPS D-brane as well, as a consequence of the fact, that this brane is embedded into supersymmetric theory. On the other hand, DBI term
 is not invariant under $\kappa$ symmetry alone, so we cannot gauge away one half of 
fermionic degrees of freedom. The same is true also for BPS D-brane, but
the  action for this brane must contain WZ term and this whole action is
invariant under $\kappa$ symmetry, leading to possibility of  gauging away one half of 
fermionic degrees of freedom, which results into BPS D-brane.

Sen also showed, how to incorporate effect of tachyon into this action. Because tachyon has mass of order of string scale, there is no systematic way how to
obtain effective action for it, but we can estimate form of this action on
some general grounds. The presence of  tachyon was incorporated into original
action for non-BPS D-brane with introducing  the some $F$ function, that has a basic 
property, that for tachyon equal its vacuum value, this function is zero, which leads to the
vanishing of action for non-BPS D-brane resulting into appearance of super
symmetric vacuum of IIA theory.

In previous paper\cite{Kluson}, we have used the action for non-BPS D-brane,
proposed by Sen in ref. \cite{Sen}, for construction of BPS brane from higher
dimensional non-BPS D-brane.  For better understanding of new results, we review
 this construction in following part.      
 
In this paper, we generalize this construction for system
of $N$ non-BPS D9-branes, in order to obtain relation of our construction with
$K$-theory construction, proposed in ref.\cite{Horava}. With using the tachyon
 kink solution given in \cite{Horava}, we obtain action, that can be expressed as sum of actions for brane and antibranes. We will argue, that this approach
is not completely correct, because we know, that this system is unstable, so 
we could expect presence of tachyon field. Then we generalize tachyon kink solution and we obtain action, that has not been previously known in literature, but we 
will argue, that this action is exactly action for system of brane and antibrane. The arguments for validity of this conjecture will then be given. Firstly, we will show, that kink solution of tachyon  on world-volume of  pair brane+antibrane results, as expect, to the unstable non-BPS D-brane in IIA theory. 
As a further check, we will show, that this action
 under $(-1)^{F_L}$ operation presented in \cite{Sen3} leads to the action of 
non-BPS D-brane with the same dimension as in IIA theory and then with further
sodding with $(-1)^{F_L}$ to the action for supper-symmetric D-brane in IIB theory with completely agreement with \cite{Sen3}.

As a last result, we present construction of vortex solution on world-volume of
D8-brane+antibrane pair resulting into stable BPS D6-brane in IIA theory, with
agreement with topological arguments presented elsewhere \cite{Sen3,Schwarz,Horava,witen}. This solution can be indication of  further research in  one 
step construction of lower dimensional branes from non-BPS D9-brane in IIA 
theory, in the same way as in \cite{Horava}.

This is brief outline of present paper. In conclusion, we return to the issue
with WZ terms, that not correctly reproduces this results and we present
some speculations about this problem.

\section{Action for non-BPS D-brane in IIA theory}
In this section we review construction of lower dimensional BPS brane 
from non-BPS D-brane, as was presented in \cite{Kluson}. The reader, who is
familiar with this paper, can skip into section devoted to the multibranes
 solutions on $N$ D9-branes in IIA theory
\footnote{We work in units $2\pi l^2_s=1$, metric has signature $(-,+,..+)$
and Majorana spinors are real and gamma matrices are pure imaginary.}.

We start this section with recapitulating basic facts about non-BPS D-branes in IIA theory, following \cite{Sen}.
Let $ \sigma_{\mu} , \ \mu=0,...p $ are world-volume co-ordinates on D-brane. Fields, living
on this D-brane arise as the lightest states from spectrum of open string ending on this D-brane. These open strings have two CP sectors \cite{Sen2}: first, with unit $ 2\times 2 $ matrix, which corresponds
to the states of open string with usual GSO projection $ (-1)^{F}\left| \psi \right>=\left| \psi \right > $, 
 where $ F$ is world-sheet fermion number and $ \left| \psi \right> $ is state from Hilbert space of
open string living on Dp-brane. The second CP sector has CP matrix  $ \sigma_1 $ and contains states
having opposite GSO projection $ (-1)^{F} \left| \psi \right>=-\left| \psi \right> $. The massless fields living
on Dp-brane are ten components of $ X^M(\sigma), M=0...9 $,  $ U(1) $ gauge field $ A(\sigma)_{\mu} $ and
32 component of fermionic field $ \theta $ transforming as Majorana spinor under transverse Lorenz
group $ SO(9,1) $. We can write $ \theta $ as sum of left handed Majorana-Weyl spinor and right handed
Majorana-Weyl spinor:
\begin{equation}
\theta=\theta_L+\theta_R ,\ \Gamma_{11}\theta_L=\theta_L, \ \Gamma_{11}\theta_R=-\theta_R
\end{equation}
All fields except $ \theta_R $ come from CP sector with identity matrix, while $\theta_R $ comes from 
sector with  $ \sigma_1 $ matrix. 

As Sen \cite{Sen} argued, action of non BPS D-brane should go to the action of BPS D-brane, when
we set $ \theta_R=0 $ (we have opposite convention that \cite{Sen}). From this reason, action
of non-BPS D-brane in \cite{Sen} has been constructed as supersymmetric  DBI action, which is manifestly
supersymmetric invariant, but has not $ \kappa $ symmetry, so we cannot gauge away one half of fermionic
degrees of freedom, so that this action describes non-BPS D-brane. In the following, we will not need the
general form of this action, because we would like to have clear physical interpretation of our result, so that we will work in static gauge:
\begin{eqnarray}\label{action}
S=-C\int d^{p+1}\sigma\sqrt{-\det(G_{\mu\nu}+F_{\mu\nu})} \nonumber \\
G_{\mu\nu}+F_{\mu\nu}=\eta_{\mu\nu}+\delta_{ij}\partial_{\mu}\phi^i \partial_{\nu}\phi^j+F_{\mu\nu} \nonumber \\
-2\overline{\theta}_L( \Gamma_{\nu}+\Gamma_i \partial_{\nu}\phi^i )\partial_{\mu}\theta_L
-2\overline{\theta}_R(\Gamma_{\mu}+\Gamma_i\partial_{\mu}\phi^i)\partial_{\nu}\theta_R + \nonumber \\
+(\overline{\theta}_L\Gamma^M\partial_{\mu}\theta_L)(\overline{\theta}_L\Gamma_M\partial_{\nu}\theta_L
+\overline{\theta}_R\Gamma_M\partial_{\nu}\theta_R)+ \nonumber \\
+(\overline{\theta}_R \Gamma^M\partial_{\nu}\theta_R)(\overline{\theta}_L\Gamma_M\partial_{\mu}\theta_L
+\overline{\theta}_R\Gamma_M\partial_{\mu}\theta_R) \nonumber \\
\end{eqnarray}
where $ F_{\mu\nu}=\partial_{\mu}A_{\nu}-\partial_{\nu}A_{\mu} $ is a field strength for gauge field and C is constant equal to the Dp-brane tension (we work in units $ l_s=1$) .

Now if we set $ \theta_R=0 $, we obtain action for BPS D-brane as in ref. \cite{Aganagic}.

The next thing is to include the effect of tachyon. In order to get some relation between tachyon condensation and
supersymmetric D-branes, we would like to have an effective action for massless field and tachyon living on world-volume of  non-BPS D-brane. This effective action should appear after integrating out all massive modes of open string ending on Dp-brane. Because tachyon  mass is of order of string scale,
there is no systematic way to obtain effective action for this field, but we can still study some general properties of 
this action. Following ref.\cite{Sen}, the effective action for non-BPS Dp-brane with tachyonic field on its
world-volume should have a form:
\begin{equation}\label{efaction} 
 S=-\int d^{p+1}\sigma\sqrt{-det(G_{\mu\nu}+F_{\mu\nu})}F(T,\partial_{\mu}T,D_{\mu}\partial_{\nu}T,...\tilde{G}^{\mu\nu}_A,
\tilde{G}^{\mu\nu}_S)+I_{WZ}
\end{equation}
where $ \tilde{G}^{\mu\nu}_{A,S} $ are symbols introduced in \cite{Witen2}, that are appropriate quantities for open string in presence of general background arising in closed string sector.
In this note, we are interested in the simplest case, where metric is flat and B field is zero, so that this quantities can be taken to be trivial.
\newpage
In the function $ F $ the constant  $ C $ is included , so that for $ T=0 , \ F=C $ and we obtain action for non-BPS D-brane (\ref{action}). This function has a important property \cite{Sen}, that for tachyon field  $ T=T_0 $, where $ T_0 $ is a vacuum value of tachyon field, this function is zero 
\footnote{In fact, Sen  argued, that in case of constant $ T $,  $ F $ reduces to the potential for tachyon and as a 
consequence of general form of potential for tachyon, this term is zero for $ T=T_0 $. In this article, we slightly change behavior
of this function, because we only demand, that in point on world-volume, where tachyon is equal to its vacuum value $ T_0 $, we
should recover supersymmetric vacuum, so that there are no fields living on non-BPS D-brane, so we have (\ref{dem}). 
 }:
\begin{equation}\label{dem}
F(T_0)=0
\end{equation}  
 We also have $ I_{WZ}=0 $ due to the fact, that WZ term has a form for non-BPS 
Dp-brane \cite{Billo}:
\begin{equation}\label{WZterm}
I_{WZ}=C\int_{p+1} C\wedge dT\wedge e^{ F+B}\wedge {\sl G}
\end{equation}
 where {\sl G} is geometrical part of WZ term, but in our case we take normal and tangent bundles to be trivial, so this term is trivial as well. We see, that for $ T= const. $, this terms is zero. When on the whole world-volume the tachyon is equal its vacuum value $ T_0 $ (Recall, that vacuum value of fields is the value, that minimise potential for this field), the action is identical zero  and we end with pure supersymmetric vacuum, which is satisfactory result, because we expect, that after tachyon condensation, non-BPS D-brane disappears and we obtain supersymmetric vacuum.

Now we are interested in nontrivial behavior of tachyon on world-volume of non-BPS D-brane. In other words, we would
like to construct kink solution for tachyon. As was explained elsewhere \cite{Sen1,Horava}, 
when we have one non-BPS D-brane, tachyon field must be real and consequently there is no coupling between tachyon and gauge field  living on world-volume of non-BPS D-brane. It can be shown \cite{Sen1,Sen2}, that potential for tachyon must
be even function of $ T $, so that we have two values of $ T $, that minimise potential for tachyon. In other words, we have two
ground(vacuum)  states for tachyon field : $ \{T_0,-T_0\} $
\footnote{To see, why tachyon is real, we must consider  general case of N non-BPS D-branes,  where tachyon lives
 in adjoin representation of $ U(N) $ gauge group. Because  U(1) has no adjoin representation,
tachyon field is real field with charge zero. This has important consequence in construction $ F$ function in (\ref{efaction}).}.

In order to construct supersymmetric D-brane as topological solution of fields living on world-volume of non-BPS D-brane,
we  expect, that  these massless field  are slowly varying, so that we can expand DBI action in
standard form, when derivatives of massless fields are small. As a result,  we obtain action for U(1) gauge field and massless scalars
$ \phi^i, i=p+1,...9 $, describing transverse fluctuations of D-brane, as in ordinary construction of BPS D-branes \cite{witen3,taylor}. There are also terms for fermionic fields $ \theta_R, \theta_L $, so that action has a form 
\begin{equation}\label{lowaction}
S=-\int d^{p+1}(F_{\mu\nu}F^{\mu\nu}+2\delta_{ij}\partial_{\mu}\phi^i\partial^{\mu}\phi^j  +F.T.)F(T,\partial T ...)+I_{WZ}
\end{equation},
where F.T. means term containing both left handed and right handed fermionic fields $ \theta_R,\theta_L $. We 
need not to know direct form of this term, we only must know, that in case of $ \theta_R=0 $ and $ T=0 $, this action 
goes into SYM action with gauge group U(1) describing BPS Dp-brane in low energy limit.

Now we would like to construct kink solution for the tachyon field. This means, that we must solve equation of motion for tachyon, which can be obtained from (\ref{lowaction}). Because tachyon is present only in $ F $ function, we can obtain equation of
motion for tachyon from this function. In the next part we must determine on some physical grounds form of this function. We 
expect, that interaction terms between tachyon and other massless fields  must be present  in this function, because
as in ref.\cite{Sen}, when we set  $T=0 $, we must obtain action for non-BPS D-brane.

We write $ F $ as
\begin{equation}\label{Fa}
F=C(V(T)+G(T,\theta_R,\theta_L))
\end{equation},
where V(T) is potential for tachyon and in G we have included kinetic term for tachyon and all interaction terms between tachyon and spinor fields and  we have used the fact, that tachyon has charge zero with respect of gauge field $ A_{\mu} $. For the same
reason there is no interaction term between tachyon and scalar field $ \phi^i $, because we know, that non-BPS Dp-brane
transform under T duality into non-BPS  $ D(p- 1) $-brane, which results into transformation of   component of gauge field in direction of T transformation into
scalar field describing position of T dual D(p-1)-brane in dual direction, and because tachyon has no charge with respect to gauge field, we can conclude, that there is also no interaction between tachyon and scalar field (We have flat metric and
B=0, so that we have no interaction between tachyon and light fields coming from closed string sector).

Now we must discuss possible form of $ G $ function. We know from previous discussion, that $ \theta_L $ comes
from CP sector with identity matrix and $ \theta_R $ comes from CP sector with $ \sigma_1 $. As we know, open
string amplitudes comes with factor containing trace of CP matrix. Interaction vertex with two fermionic lines
and   $ k $ tachyon lines contain following trace (we are interested only in terms with two fermionic
fields without derivatives, because interaction with more fermionic fields and with more derivatives do not give new information):
\begin{equation}
\tr((\sigma_1)^k {\bf1}\sigma_1)
\end{equation}

where first matrix corresponds to the tachyon, the second matrix to $\theta_L$ and the third matrix to  $ \theta_R $.
From the identity $ \tr(\sigma_1^{2n+1})=0 $ it is clear, that $ k $ must be odd.  We than propose interaction between tachyon field and one spinor field from left handed sector and one field from right handed sector:
\begin{equation}\label{G}
G(T,\theta_R,\theta_L)=f(T)\overline{\theta}_R\theta_L +\partial_{\mu}T\partial^{\mu}T
\end{equation}
 where $ f(T) $ are odd function of  $T$ and where we have included only terms with two derivatives of tachyon into G
\footnote{We can also consider interaction in the form: $g(T)(\overline{\theta}_R\theta_R+\overline{\theta}_L\theta_L)$, where
now g(T) is even function in $T$. As will be explained letter, it is natural to consider this term to be zero.}.

Now, we are ready to construct kink solution on non-BPS D-brane. We suppose, that tachyon is function only one co-ordinate, say
$ x_1 $.  Firstly, we should ask what happen, when
T is equal one of vacuum expectation value in one particular point. We know, that $ F $ in (\ref{lowaction})
must be zero in this point, because we should obtain pure supersymmetric vacuum  in this point:
\begin{equation}\label{F}
F=C(V(T_0)+f(T_0)\overline{\theta}_R\theta_L)=0
\end{equation},
because $ V(T_0) $ is zero and there is no reason why $ f(T_0)   $ should be zero (for example, in the simplest case
$ f(T)=T $, this function is clearly nonzero in $ T=T_0 $ ), the only possibility is to take either $ \theta_R $ or
$ \theta_L $ to be zero. We choose $ \theta_R=0 $ in the point, where $ T=T_0 $. We must also mention, that other fields are not restricted in this particular point.

In  kink solution,  tachyon approaches in limit 
$ x_1 \rightarrow -\infty $ one vacuum value (say $ -T_0 $) and then in limit $ x_1 \rightarrow \infty $  the vacuum
value $ T=T_0 $. We can take special limit of this solution, when tachyon is different from its vacuum expectation 
value $ T_0 $ only in small region around point $ x_1=0 $ and in extreme case, $ T $ has a form:
\begin{equation}\label{T}
T(x_1)=\left \{ \begin{array}{c} -T_0,\quad x_1 < 0 \\
                                      T_0 \quad x_1 > 0 \\
\end{array}\right.
\end{equation}
Of course, this is very special function and there could be some opinions, why we do not take terms with
higher derivatives into $ G $, but we  expect,  that these terms make  behaviour of our solution more singular, so that in any case,
$ F  $ looks like $ F \sim \delta(x_1) $(this can be seen from the fact, that potential term in (\ref{F}) is finite function
and singular behaviour comes from derivative term). Now
from (\ref{G}) we can obtain the equation of motion for tachyon
\begin{equation}
\frac{d^2T}{dx^2_1}+\frac{df}{dT}(\overline{\theta}_R\theta_L)+\frac{dV}{dT}=0
\end{equation}
We know, that $ \frac{dV(T)}{dT}  \sim \delta(x_1) $ and $ \frac{dT^2}{dx^2_1} $ is singular as well, but because delta function is even
function, this function is odd,  we can expect, that these two terms cancel among themselves, and because we suppose,
that $ f(T)  $ is polynomial  in  $ T $, that is finite in region $ x_1=0 $, we can conclude, that in order to satisfy equation of
motion for tachyon, we must demand, that  $ \theta_R $ is zero  in $ x_1=0 $ as well. 
\begin{equation}\label{theta}
\theta_R=0
\end{equation}

This result is with agreement with our original assumption, that massless fields vary slowly on world-volume of non-BPS D-brane,
because in case of nontrivial  dependence  of $ \theta_R $ on $ x_1 $,  its derivative would by proportional
 to the  delta function and then $ \theta_R $ certainly does not vary slowly on world-volume of non-BPS D-brane.

We know, that in extreme case (\ref{T}), we have
\begin{equation}\label{F1}
F(x_1) \sim C\delta(x_1)
\end{equation}
\footnote{We give more precise numerical factor bellow, after fixing vacuum value of tachyon.}
and when we insert this term into (\ref{lowaction}), we obtain integral over  (p-1) spatial direction, which can suggest, that we have obtained  low energy action for BPS D(p-1)-brane (Condition $ \theta_R=0 $ in original action leads to the action for BPS Dp-brane)
 \footnote{we must remember, that we 
are working in units $ 2\pi \l_s=1 $, so that tension of Dp is ( $ C \sim 1/g $ ), but when we would work in units, where 
$ 2\pi l_s $ is not one , than integral over $ x_1 $ would be multiplied with $ l_s $ , so we would obtain tension for D(p-1)-brane as
$ T \sim Cl_s \sim \frac{l_s}{gl_s^{+1}} \sim \frac{1}{gl_s^p} $, which is a correct tension for D(p-1)-brane.}, but there are two puzzles,
that must be resolved. Firstly, when we insert (\ref{F1}) into (\ref{lowaction}), we also obtain following terms
\[ \partial_{x_1} \Psi |_{x_1=0} \],
where $ \Psi $ is any massless world-volume field. But we know, that there is no dependence of any field in resulting (p-1)
dimensional action on $ x_1 $, so we must demand, that this term is zero.
\begin{equation}
\partial_{x_1}\Psi |_{x_1=0}=0
\end{equation}

The other problem arises with scalars describing transverse fluctuation of D-brane.
Any Dp-brane has (9-p) world-volume scalars describing fluctuation of this D-brane in transverse
directions. Original non-BPS action  has (9-p) scalar fields, but now  we would like to have action for D(p-1)-brane, so that there
should  be (10-p) scalars. We can obtain this additional scalar from gauge field. Remember, that in original action we
have had a term (now M, N goes from 0...p and $ \mu,\nu =0...p-1 $ ):
\begin{equation}
 F_{MN}F^{MN}=2F_{1\mu}F^{1\mu}+F_{\mu\nu}F^{\mu\nu} 
\end{equation}

We see, that additional scalar can be first component of gauge field $ A_1$, that has kinetic term
\begin{equation}
2F_{1\mu}F^{1\mu}=2\partial_{\mu}A_1\partial^{\mu}A^1
\end{equation},
that has the same form as term for scalar fields in (\ref{lowaction}), so  that after renaming $ A^1=~\phi^p $, we obtain
(10-p) scalars, the correct number of scalar fields describing transverse fluctuations for D(p-1)-brane. From previous facts and from  (\ref{theta})  we conclude, that 
kink solution on non-BPS Dp-brane forms BPS D(p-1)-brane, that is localised in $ x_1 $ direction. This can
be also confirmed with  WZ term \cite{Billo}, that in extreme case has a form ($ dT(x_1)=\delta(x_1)dx_1 $)
\begin{equation}
I_{WZ}=\int_{R^{1,p-1}} C\wedge e^{ F}
\end{equation}
which has the same form as WZ term for BPS D(p-1)-brane.
\section{Multibrane solutions on $N$ 9-branes in IIA theory}
In this part we would like to analyse tachyon solution on $N$ 9-branes in IIA theory
proposed by Ho\v{r}ava in ref.\cite{Horava}. Firstly, we must say a few words
 about potential for tachyon taking value in adjoin representation of 
 $U(N)$. We can expect this  potential  in the form:
\begin{equation}
V(T)=-m^2\tr(T^2)+\lambda\tr(T^4)
\end{equation}
This potential leads to the ground state value of tachyon, where all eigenvalues are equal, possibly up to sign.  
\footnote{This potential is not zero for $T=T_0$.To get potential, that is zero for $T=T_0$. we add constant to the 
 potential, that leads to   $V(T)=-m^2\tr(T^2)+\lambda\tr(T^4)+\lambda\tr(T_0)^4$. This function is zero for $T=T_0$. In the following, when we  speak about
 potential for tachyon, we will mean this expression.}.

As in ref. \cite{Horava}, we  take tachyon vacuum value in the form:
\begin{equation}\label{sola}
T_0=T_v\left(\begin{array}{cc}1_{N-k} & 0 \\
                                             0 & 1_k \\
            \end{array}\right)
\end{equation}
We see, that this vacuum value is invariant under gauge transformations of the form $U(N-k)\times U(k)$,because each
element, that belongs to this gauge group, has a form 
\[ \left(\begin{array}{cc} A & 0 \\
                                    0 & B \\ \end{array}\right) \]
, where $A \in U(N-k) $ and $B \in U(k)$.
Now we  take solution of equation of motion, where tachyon forms kink. This 
kink solution should lead to the  collection of 8-branes and antibranes
\cite{Horava}. In particular, tachyon can look around point $x=0$ as:

\begin{equation}\label{sol}
T=\left(\begin{array}{cc} x 1_{N-k} & 0 \\
                                     0 & -x1_{k} \\ \end{array}\right)
\end{equation}

This kink form state with $N-k$ 8-branes in point $x=0$ and $k$ anti 8-branes
 in the  same point . Of course, we can expect, that this configuration is unstable, which can be seen from supergravity point of view
as a result of attractive force between brane and antibrane, so that $k$ branes anhilate with $k$ antibranes resulting into 
stable configuration of $N-2k$ branes (We consider situation, where $N-k$ is 
larger than $k$.). Firstly, we will analyse the tachyon solution   (\ref{sol}) in the same way as in case of one non-BPS D-brane presented in the 
previous section and we  show,
that this analysis leads to the action describing $N-k$ D8-branes and $k$
 anti D8-branes. However we will see, that  tachyon field in the form (\ref{sol}) leads to the solution, where is not complete clear instability of these configuration. In other words, in
resulting action the terms describing tachyon arising from open string connecting brane and antibrane are missing. Then we show, that
simple modification of tachyon solution in (\ref{sol}) leads to the expression,
that can be interpreted as correct action for the configuration of  branes and antibranes.

We start our analysis with WZ term for non-BPS D-brane \cite{Billo}, that has
 a form:
\begin{equation}
S=a\int C\wedge d\left(\tr T e^{F}\right)
\end{equation}

where $a$ is a constant proportional to D-brane charge and where geometrical
 part is trivial due to the fact, that there is no normal gauge bundle.
 We see, that this term is gauge invariant, because contain trace over
 matrices in adjoin representation. But when we make expansion of exterior
 derivate  \[\tr d(Te^F)=\tr dT\wedge e^F+\tr T d(e^F)\]
we would like to have each term invariant separately. Natural generalisation is to take covariant derivate in place of ordinary exterior derivate, so 
we obtain the result:
\begin{equation}\label{newWZ}
I_{WZ}=\int C_{p+1} \wedge \tr D(Te^F)
\end{equation}
where $ D=d +[A, ] $
\footnote{ Arguments for the using of covariant  derivate in place of  ordinary  derivate come from T-duality. We consider  term $\tr YT$, where $Y$ is a field on 
 D-brane describing transverse fluctuation of brane in this direction. When we
apply T-duality in this direction, we obtain term $\tr DT$, where scalar field
was replaced with covariant derivate. We can than
expect, that all derivates should be replaced with covariant derivates.}. 
As a check, we would see, that this new WZ term correctly reproduce charge of
D8-branes and D8-antibranes,  after using (\ref{sol})
\footnote{There are important objections to this result, that will be           presented in conclusion.}. 

We have conjectured, that tachyon field $T$ describing configuration of $N-k$ 
D8-branes and $k$ D8-antibranes has  a 
form (\ref{sol})(As in \cite{Horava} we do not consider possible convergence    factor.). We  suppose, that near the core of the vortex,
tachyon is in its vacuum value. We have also seen, that this $T$ breaks gauge
 symmetry into $U(N-k)\times U(k)$. We would
like to know, whether (\ref{sola}) places some restrictions on gauge fields. 
We demand, that this configuration must solve equation of motion for tachyon,
 so that  exterior derivate  $DT$ is zero in the point, where tachyon is equal to
its vacuum value. Because except the point $x=0$, tachyon is in its 
 vacuum value (we consider solution in the form (\ref{T}).), we have $ dT=0$,
 so that  we obtain condition for gauge field outside  the point $x=0$:
\begin{equation}\label{con}
[T_0,A_x] =0
\end{equation}
We write $A_x$ as follows:
\[ A_x=\left( \begin{array}{cc} A & B \\
                              C & D \\ \end{array}\right) \] and after inserting into (\ref{con}), we obtain:
\begin{equation}
\left[ \left(\begin{array}{cc} 1_{N-k} & 0 \\
                                         0 & 1_k \\ \end{array}\right),\left(\begin{array}{cc} A & B \\
                                                                                                                    C & D \\ \end{array}\right) \right]=
\left(\begin{array}{cc} 0 & 2B \\
                                 -2C & 0 \\ \end{array}\right)
\end{equation} 
, where we omit constant factor. We than see, that $B,C=0 $, in other words, gauge fields divide into two parts corresponding to the gauge group
$U(N-k)$ and $U(k)$ respectively. When we look at (\ref{sol}), we see, that tachyon field breaks gauge symmetry also in point $x=0$, we
can than expect, that gauge field is in two subgroups of gauge group everywhere as well, because we do not expect singular
change in the gauge group. This result has a direct consequence, because we 
have 
\begin{equation}
[T(x),A_x(x)]=0
\end{equation}
for all $x$, so that covariant derivate $T$ is the same as ordinary derivate
: $ DT=dT$. We than obtain
\begin{equation}
\tr D(T e^F)=\tr DTe^F+ \tr T D(e^F)=\tr dTe^F 
\end{equation}
where we have used the Bianchi identity $DF=0$.
From (\ref{T}), that holds for all diagonal elements, we obtain:
\begin{equation}
dT(x)=\left(\begin{array}{cc} \delta(x) 1_{N-k} & 0 \\
                                            0 & -\delta(x)1_k \\ \end{array}\right)dx
\end{equation}
and using the know result from algebra:
\begin{equation}
\tr(\left(\begin{array}{cc} A & 0 \\
                               0  & B \\ \end{array}\right)\left(\begin{array}{cc} C & 0 \\
                                                                                                       0 & D \\ \end{array}\right)=
=\tr(AB)+\tr(CD)
\end{equation}
we arrive to the final result :
\begin{eqnarray}\label{WZ2}
I=\int C\wedge \tr dT \wedge e^F=\int C\wedge dx \delta(x) \tr_{N-k}e^F -
\int C\wedge dx \delta(x) \tr_k e^{F'}= \nonumber \\
=\int_{D8} C\wedge \tr_{N-k}e^F -\int_{\overline{D8}}C \wedge \tr_k e^{F'} \nonumber \\
\end{eqnarray}
where $F \in U(N-k) $ is identified with field strength for $N-k$ 8-branes and $ F' \in U(k) $ is field
strength identified with $k$ anti-8-branes.  

This result can be expected in the action for $N-k$ branes and $k$ antibranes, consequently can serve as a justification of validity of construction branes 
through tachyon condensation. But this cannot be final result, because there
is missing term expressing instability of this configuration, as in
\cite{Wilkinson}. We will see, that we do not obtain correct result even in
 case of more general tachyonic kink solution. We return to this important 
question in conclusion.

As a next step, we start analyse action for non-BPS D-brane in the same way 
as in previous section. Because we do not know exactly the
DBI action for system of N-brazes, we take low energy limits with all derivatives and all field strength of massless fields
to be small, so that generalised action for $N$ 9-branes looks like:
\begin{eqnarray}\label{act2}
S=-\int(1+\frac{1}{4}) (\tr F_{MN}F^{MN}+2i \tr \theta_L\Gamma^MD_M\theta_L+ \nonumber \\
+2i \tr\theta_R\Gamma^MD_M\theta_R)F(T,DT,...)+I_{WZ} \nonumber \\
\end{eqnarray}
Where we have introduced a constant term and  as in ref.\cite{Sen} we expect, that for $T=0$ this action comes to the action for
$N$ non-BPS D-branes, so that we have $F(T=0)=C$. From this reason, the terms in brackets must
be separately gauge invariant and $F$ must be gauge invariant as well.  
 We generalise results from ref.\cite{Kluson} for the case of $N$ non-BPS D-branes, so that we  we can expect $F$ in the form:
\begin{equation}\label{F2}
F(T,DT..)=\tr D_MTD^MT+\tr (f(T)\overline{\theta_R} \theta_L) +V(T)
\end{equation}
 Now we suppose kink solution of the tachyon field in the form (\ref{sol}) 
(More precisely, we take kink solution with diagonal term to be equal to 
(\ref{T}).). We know, that this solution
break gauge symmetry into $ U(N-k)\times U(k) $. Again we demand, that in some particular point $x$ where $ T=T_0$ this system
is equivalent to the supersymmetric vacuum, so we have $ F(T_0)=0$ and this condition place some constrains on
massless fields
\footnote{This condition is equivalent to the condition, that tachyon must
 obey equation of motion $\frac{\delta F(T)}{\delta T}=0$.}
. Firstly, we analyse gauge fields.

In place, where $ T=T_0 $ , we must have $DT=0 $, which again leads to the separating gauge fields  into
two parts, $A_x$, transforming in adjoin representation of $U(N-k)$ and $A'_x $ transforming in $U(k)$. Even if
the  breaking of gauge group  appears only in x-th component, this must hold
 for  the other components as well, because the braking of gauge group does 
not depend on component of particular field. We see, that there are no other conditions for these fields, except
the well know fact, that must vary slowly. From the previous section we know,
that massless fields must be independent on $x$ direction for tachyon field 
forming kink solution. With using this fact, we obtain from kinetic term for gauge field following result:
\begin{equation}\label{gauge1}
\tr F_{MN}F^{MN}=\tr_{N-k}F_{\mu\nu}F^{\mu\nu}+\tr_k F'_{\mu\nu}F^{'\mu\nu}+\tr_{N-k}F_{\mu x}F^{\mu x}
+\tr_k F'_{\mu x}F^{'\mu x}
\end{equation}
where: $M,N=0,...9 , \ \mu,\nu=0,...8 $.
We see, that the first two terms correspond to the field strengths of gauge fields living on world-volume of 8-branes and
8-antibranes, and the third and the fourth term correspond to the transverse fluctuation of 8-branes and antibranes after renaming
$ A_x=X$:
\begin{equation}\label{gauge2}
\tr_{N-k}F_{\mu x}F^{\mu x}=2\tr_{N-k}(\partial X +[A_{\mu},X])^2=2\tr_{N-k}(D_{\mu}XD^{\mu}X)
\end{equation}
and the same form for  antibranes.

In the same way we will analyse  the   fermionic term in point, where $T=T_0 $. In this point, the function $f(T)$ in 
(\ref{F2}) looks like:
\begin{equation}
f(T_0)=\left(\begin{array}{cc} K(T_0)1_{N-k} & 0 \\
                                            0          & K(T_0)1_k \\ \end{array}\right)
\end{equation}
where function $K$ follows from the form of $f$. Now we write fermionic terms 
as
\begin{equation}\label{fermi}
\theta_R=\left(\begin{array}{cc} A & B \\ 
                                                C & D \\ \end{array}\right) , \ \theta_L=\left(\begin{array}{cc} E & F \\
                                                                                                                                           G & H \\ \end{array}\right)
\end{equation}
where $ A,E $ are matrices with $ (N-k)$ rows and $ (N-k)$ columns, $D,H$ are $ k\times k$ matrices and 
$C,G$ are $ k\times(N-k)$ matrices and finally $B,F$ are matrices with $(N-k)$ rows and $k$ columns.
With this expansion, the potential term in point $T=T_0 $ is
\begin{equation}
\tr f(T)\overline{\theta_R}\theta_R
=\tr_{N-k} K(T_0)(\overline{A}E+\overline{B}G)+\tr_k L(T_0)(\overline{C}F+\overline{D}H)
\end{equation}
Again, this term must vanish. As in case of single non-BPS D-brane, we have some possibilities to obey this
condition and we take 
\begin{equation}
A=0, \ D=0
\end{equation}
in point $T=T_0$, but due to the same arguments as in one non-BPS brane, this must be extended over the whole
axis $x$. The other choices are either $B,C=0$ or $G,H=0$. We take $G,H=0$
, in order to obtain nontrivial interaction, when we will consider generalised
tachyon solution in the next section.
\begin{equation}
G=0, \ F=0
\end{equation}

In the same arguments as before, these two matrices must be zero on the whole $x$ axis. 
As a next step,  we insert this result into kinetic terms for fermions in (\ref{act2}). We start with $\theta_R$. As in 
previous situation for gauge field, we demand, that fermionic fields are independent on $x$ co-ordinates:
\begin{equation}\label{con2}
\partial_x \theta_L=0, \partial_x \theta_R=0
\end{equation}
so that derivative part in covariant derivate for $\theta_R$ has a form:
\begin{eqnarray}
\tr \overline{\theta_R}\Gamma^M\partial_M\theta_R=\tr_{N-k}\overline{C}\Gamma^{\mu}\partial_{\mu}C+\tr_k
\overline{B}\Gamma^{\mu}\partial_{\mu}B 
\end{eqnarray}
and term with gauge field is:
\begin{eqnarray}\label{spin3}
\tr \overline{\theta}_L\Gamma^M[A_M,\theta_L]=\tr_{N-k}\overline{C}\Gamma^{\mu}(A'_{\mu}C-CA_{\mu})
+\tr_k\overline{B}\Gamma^{\mu}(A_{\mu}B-BA'_{\mu})+\nonumber \\
+\tr_{N-k}\overline{C}\Gamma^9(X'C-CX)+\tr_k\overline{B}\Gamma^9(XB-BX') \nonumber \\
\end{eqnarray}
where we now take $x$ direction as 9-th direction and we have renamed $A_9$ as $X$ and $\mu=0,...8$. We can see, that the last term expresses nonlocal character of spinor fields, because relates scalar components living on 
brane with scalar component living on antibrane. It seams to be strange and we return to this puzzle in a moment.
We must also argue, that appearance of these terms is natural, again from the reason of T-duality, where from
standard gauge covariant term, that will be written bellow, we obtain exactly this term.

The gauge covariant terms for $C,B$ are:
\begin{eqnarray}\label{spin}
\tr_{k}\overline{B}\Gamma^{\mu}(\partial_{\mu}B+(A_{\mu}B-BA'_{\mu})) \nonumber \\
\tr_{N-k}\overline{C}\Gamma^{\mu}(\partial_{\mu}C+(A'_{\mu}C-CA_{\mu})) \nonumber \\
\end{eqnarray}

In the following part, we would like to argue, that this Lagrangian correspond to the fermionic degrees
of freedom arising from string theory point of view as a fermionic states from R sector of open string
stretching between $N-k$ 8-branes and $k$ 8-antibranes, so that from string theory we have transformation
law for these spinors 
\begin{equation}\label{trsp}
 \psi \rightarrow g\psi h^{-1} , \ g\in U(N-k), \ h \in U(k) 
\end{equation}
\footnote{This transformation law is the same as for tachyon, that was presented in \cite{witen},
because we have from string theory analysis of Chen-Paton factors:
$ \psi^{'ij}=g^{ik}h^{*jl}\psi^{kl}=
g^{ik}(h^{\dag})^{lj}\psi^{kl}=(g\psi h^{-1})^{ij} $ .}. We must also note, that these spinors, are complex,
where this complex property is a consequence of  the fact, that these fermions come from the oriented string, so that
we have two degrees of freedom for each string starting on i-th brane and ending on j-th brane (In case of
non oriented strings theory, these fermions will be real, because they are Majorana-Weyl spinors and can be
real in appropriate basis for Dirac matrices.). 
From the transformation property of spinor modes on $N$ 9-branes under transformation of the form 
\[U=\left(\begin{array}{cc} g & 0 \\
                                        0 & h \\ \end{array}\right) , g \in U(N-k), h\in U(k) , U \in U(N-k)\times U(k) \]
and
\[ \theta \rightarrow U\theta U^{-1} \]
that holds for both left and right handed spinors, we obtain for 
\[ \theta_R=\left(\begin{array}{cc} 0 & B \\
                                                   C & 0 \\ \end{array}\right) \]
the transformation property for $C$ and $B$ 
\begin{equation}
\tilde{B}=gBh^{-1}, \tilde{C}=hCg^{-1}
\end{equation}
We see, that these spinors transforms as in (\ref{trsp}), so we can interpret  these spinors as spinor states  coming in string theory point of view from sector of strings connecting
brane and antibrane. As other check, we can show, that (\ref{spin}) is invariant under $U(N-k)\times U(k) $
symmetry, where gauge fields transforms as:
\begin{equation}
\tilde{A}= -dgg^{-1}+gAg^{-1} , \tilde{A'}=-dhh^{-1}+hgh^{-1}
\end{equation}
where $g, h $ have the same meaning as before.
We introduce  covariant derivative 
$\hat{D}B=dB+AB-BA'$, that transforms as:
\begin{equation}
\tilde{\hat{D}B}=g\hat{D}h^{-1}
\end{equation}
so that first term in (\ref{spin}) is gauge invariant:
\begin{equation}
\tr_{k}\tilde{\overline{B}}\Gamma\tilde{\hat{D}}B=\tr_{k}(h\overline{B}g^{-1}g\Gamma\hat{D}Bh^{-1})=
\tr_{k}\overline{}B\Gamma\hat{D}B
\end{equation}
where we have used($ \tilde{B}=gBh^{\dag} \Rightarrow \tilde{\overline{B}}=\tilde{B^{\dag}}\Gamma^0=g\overline{B}h^{-1} $)
and where for simplicity write $\Gamma D $ instead $ \Gamma^{\mu}D_{\mu}$.
\footnote{ We must say a few words about notation. When we write $\tr\overline{\psi}\psi=\tr \psi^{T\dag}\Gamma^0\psi
=\sum_{i,j}\psi^{*T}_{ij}\Gamma^0\psi_{ij} $, where symbol $T$ over $\psi$ means operation of transposition related to the
spinor indexes as in ordinary QFT and indexes $i,j $ can be interpreted as labeling different species in internal symmetry.} 
From the fact, that $ \theta=\theta^{\dag}$, we obtain $C^{\dag}=B$, so that both two terms in (\ref{spin}) are the same.

Now we would like to analyse form of $\theta_L$. We see, that in (\ref{fermi}), $F$ and $G$ are zero. We than have:
\begin{equation}\label{g1}
\overline{\theta_L}\Gamma^M\partial_M\theta_L=\tr_{N-k}\overline{E}\Gamma^{\mu}\partial_{\mu}E
+\tr_k\overline{H}\Gamma^{\mu}\partial_{\mu}H
\end{equation}
where we again use (\ref{con2}). The part with gauge fields gives:
\begin{eqnarray}\label{g2}
\overline{\theta_L}\Gamma^M[A_M.\theta_L]=\tr_{N-k}\overline{E}\Gamma^{\mu}[A_{\mu},E]+
\tr_k\overline{H}\Gamma^{\mu}[A'_{\mu},H]+ \nonumber \\
+\tr_{N-k}\overline{E}\Gamma^9[X,E]+\tr_k\overline{E}\Gamma^9[X,H] \nonumber \\
\end{eqnarray}
.(We rewrite $ E \rightarrow \theta, H \rightarrow \theta'$).

$\theta$ is field coming from the string sector living on $N-k$ 8-branes and   $\theta'$ 
is field coming from 
  the string sector living on $k$ anti 8-branes. With this notation we obtain from (\ref{g1},\ref{g2}) the well known
terms for spinor fields living on branes:
\begin{eqnarray}\label{spin2}
 \tr_{N-k}\overline{\theta}(\Gamma^{\mu}D_{\mu}\theta+\Gamma^9[X,\theta]) \nonumber \\
\tr_k\overline{\theta'}(\Gamma^{\mu}D_{\mu}\theta'+\Gamma^9[X',\theta']) \nonumber \\
\end{eqnarray}

We are now ready to write the complete action after tachyon condensation. With using (\ref{WZ2},\ref{gauge1},\ref{gauge2},
\ref{spin3},\ref{spin},\ref{spin2}) and with the fact, that as in case of one non-BPS brane, we have $F(x)=C\delta(x)$
\footnote{Again, we do not carry about precise numerical factors, which will be
given in the next section.}
, so we obtain
\begin{equation}\label{action3}
S=S_{brane}+S_{antibrane}
\end{equation}
\begin{eqnarray}\label{actbrane}
S_{brane}=-C\int_{R^{1,8}}\tr_{N-k}\left( F_{\mu\nu}F^{\mu\nu}+2D_{\mu}XD^{\mu}X+2i
\overline{\theta}(\Gamma^{\mu}D_{\mu}\theta+\right.\nonumber \\
\left.+\Gamma^9[X,\theta])\right) 
+\tr_{k}(\overline{B}\Gamma^{\mu}\tilde{D}_{\mu}B+\overline{B}\Gamma^9(XB-BX'))\nonumber \\
+\int_{R^{1,8}}C\wedge \tr_{N-k}e^F \nonumber \\
\end{eqnarray}
and for anti-brane
\begin{eqnarray}\label{actantibrane}
S_{anti}=-C\int_{R^{1,8}}\tr_{k}\left( F'_{\mu\nu}F'^{\mu\nu}+2D_{\mu}X'D'^{\mu}X'+
2i\overline{\theta'}(\Gamma^{\mu}D'_{\mu}\theta'+\right.\nonumber \\
\left.+\Gamma^9[X',\theta'])\right) 
+\tr_{k}(\overline{B}\Gamma^{\mu}\tilde{D}_{\mu}B+\overline{B}\Gamma^9(XB-BX'))\nonumber \\
-\int_{R^{1,8}}C\wedge \tr_{k}e^{F'} \nonumber \\
\end{eqnarray}

We see, that due to the tachyon condensation into kink solution of the form (\ref{sol}), we obtain action for
$N-k$ 8-branes and $k$ 8-branes. We also see, that both action for branes and antibranes contain the common
term, that from the string theory point of view comes from R-sector of string connecting brane and antibrane. The
form of the whole action, that is expressed as sum the two action for different branes and antibranes suggests, that
from the tachyon solution, we obtain brane and antibrane, that are \emph{distinguishable}, but there is a difference from
ref.\cite{Pesando}. In that paper,a covariant derivate for tachyon in the
action for brane, that contain only gauge field living
on brane, is $DT=dT+AT$ and on antibrane is $D'T=dT-A'T$,  while in case of 
\emph{indistinguishable} branes, we have
covariant derivate for tachyon $DT=dT+AT-A'T$ (In our case, tachyon correspond to the spinor field coming from
string connecting brane and antibrane.). In our approach, we have branes and antibranes, that seam to be distinguishable,
which can be natural property of them, because branes carry RR charge and antibranes carry opposite charge, but covariant
derivate for spinor field contain both gauge field living on brane and antibrane:$\tilde{D}B=dB+AB-BA' $. It seams to us,
that this covariant derivate is more appropriate for field, that arise from string sector connecting brane and antibrane, because
this field, that for example now live on brane, can also field gauge field living on antibrane. In (\ref{action3}) is also term,
that relates transverse fluctuation of brane and antibrane due to the spinor field $ (XB-X'B)$, that again supports idea, that
these field can have string theory interpretation in terms of string connecting brane and antibrane. 

The other puzzle is with tachyon. We know, that system of brane and antibrane is unstable, so that there should be tachyon presented in the action reflecting instability of these configuration. In the next section we will show, that simple modification of 
tachyon solution in (\ref{sol}) leads to correct result.
\section{Generalised tachyonic solution}

Firstly we must ask, what form of tachyon field we would like. We demand, that outside the point $x^9=0$, where branes and
antibranes are sitting, the system should be equivalent to the supersymmetric vacuum, so that tachyon must be in their vacuum
value 
\begin{equation}
T_0=T_v\left(\begin{array}{cc}1_{N-k} & 0 \\
                                             0 & 1_k \\
            \end{array}\right)
\end{equation}
The question is, how this solution looks like around point $x^9=0$. Again we would like to have kink solution on diagonal elements
of tachyon fields, that in extreme case is
\begin{equation}
T(x^9)_{diag}=T_v\left(\begin{array}{cc}-1_{N-k} & 0 \\
                                             0 & 1_k \\
            \end{array}\right) , x^9<0 ,\ T(x^9)_{diag}=T_v\left(\begin{array}{cc}1_{N-k} & 0 \\
                                                                                                       0 & -1_k \\ \end{array}\right) , x^9>0
\end{equation}
but generally, there could be nonzero tachyon field in non-diagonal form in point $x^9=0$, which must be zero
in points outside $x^9=0$, so that we take solution in the form 
\begin{equation}\label{sol2}
T(x)=\left(\begin{array}{cc} T_0(x)1_{N-k} & T(y)\delta(x) \\
                                         \overline{T}(y)\delta(x) & -T_0(x) 1_k \\ \end{array}\right)
\end{equation}
where $T_0(x)$ is given in (\ref{T}) and $y$ mens co-ordinates on $R^{1,8}$. We insert this function into (\ref{F2}) and we 
will analyse various constraints as in previous case. The analysis is almost the same as before with difference in point $x=0$, where new phenomena arise. As in previous part, in point, where $T=T_0$, the system must be
equivalent to the supersymmetric vacuum, which again leads to the breaking of gauge field as in form (\ref{con}). This leads to following results:
\begin{equation}\label{Dx}
\tr D_xTD_xT=\tr_NT^2_v\delta^2(x)+2(T(y)\frac{d\delta(x)}{dx}+(A_9T-TA'_9)\delta(x))^2
\end{equation}
where we use of gauge fields in the form 
\begin{equation}
A_M=\left(\begin{array}{cc} A_M & 0 \\
                                          0 & A'_M \\ \end{array} \right)
\end{equation}
where $ A_9 \rightarrow X $. Now as in ordinary QFT, we can take $\delta^2(x)\sim \delta(x) $ and in the
same way $ (\frac{d\delta(x)}{dx})^2 \sim \frac{d\delta(x)}{dx} $. Then in the second term in (\ref{Dx}) are expressions, that are proportional to $ \frac{d\delta(x)}{dx} $  leading after putting into integral to the results:
\[\int(....)\frac{d\delta(x)}{dx}dx=-\int \frac{d(...)}{dx}\delta(x)=-\frac{d(...)}{dx}|_{x=0} \]
but from the requirement, that derivation of any massless field in point $x=0$ is zero, we see, that contribution of these terms is zero( in the following, we will write $x$ instead $x^9$, only in situation, where gamma matrices will be presented, we use $x^9$.).
As a result, we obtain from (\ref{Dx}):
\begin{equation}
(NT_v+2\tr_{N-k}(XT-TX')(X'\overline{T}-\overline{T}X))\delta(x)
\end{equation}
where for simplicity $T$  means $T(y)$.
In the same way, we obtain:
\begin{equation}
D_{\mu}T(x,y)=\left(\begin{array}{cc} 0 & \tilde{D}_{\mu}T \\
                                                       \overline{\tilde{D}_{\mu}T} & 0 \\ \end{array}\right)\delta(x)
\end{equation}
where we have defined $ \tilde{DT}=dT+AT-TA' , \overline{\tilde{D}T}=d\overline{T}+A'\overline{T}-\overline{T}A $.
As a result, we obtain 
\begin{equation}\label{Dy}
\tr DT_{\mu}DT^{\mu}=2\tr_{N-k}(\tilde{D}T^{\mu}\overline{\tilde{D}_{\mu}T})\delta(x)
\end{equation}
Now we comes to the analysis of potential.  This potential has form:
\begin{equation}\label{pot}
V(T)=-m^2\tr(T^2)+\lambda\tr(T^4)+\lambda\tr(T_0)^4
\end{equation}
having property $V(T_0)=0$. We see, that in points $x\neq 0$, this potential is zero. For points around $x=0$, we must return to the (\ref{T}), where we now must remember, that this solution is express as limit of function $cx$, where c goes to infinity. This means, that in point $x=0$, we have $T_0=0$.(Remember, that this same is true in function $f(T)$ in interaction between fermions and tachyon, but we demand continuous functions of massless fields, so when $\theta_R$ is zero in all point except $x=0$, is zero  in point $x=0$ as well). We see, that in point $x=0$ the solution in(\ref{sol2}) is :
\begin{equation}\label{sol3} 
T(x=0)=\left(\begin{array}{cc} 0 & T(y) \\
                                      \overline{T}(y) & 0 \\ \end{array}\right)
\end{equation}
(we must remember, that this solution is formally multiplied $\delta(0)$. When we insert (\ref{sol3}) into (\ref{pot}), we obtain:
\begin{equation}\label{pot2}
V=-2(-m^2\tr_{N-k}(T\overline{T})+\lambda \tr_{N-k}(T\overline{T})^2)\delta(x)
\end{equation}
 We have multiplied this potential with delta function, because in all points except $x=0$ this is zero and in point $x=0$ there is a delta function coming from non-diagonal terms in (\ref{sol2}). From this reason we can neglect constant term in (\ref{pot}), because
this is finite. 

As a last step we arrive to the terms with fermions in (\ref{F2}). As in previous section,  for  $T=T_0$ we obtain, that diagonal
terms in $\theta_R$ are zero: $A, D=0$,  and in $\theta_L$ the non-diagonal terms are zero: $ F, G=0 $. By requirement of continuously, these fields are zero in point $x=0$ as well. As a result, after inserting (\ref{sol3}) into (\ref{F2})   we obtain for fermionic terms in the point  
$x=0$:
\begin{eqnarray}\label{fermi2}
\tr f(T)\overline{\theta_R}\theta_L=\left(\begin{array}{cc} 0 & f(T) \\
                                                                              f(\overline{T}) & 0 \\ \end{array}\right)
\left(\begin{array}{cc} 0 &\overline{C} \\
                          \overline{B} & 0 \\ \end{array}\right)
\left(\begin{array}{cc} E & 0 \\
                                 0 & G \\ \end{array}\right)= \nonumber \\
=\left(\tr_{N-k}(f(T)\overline{B}E)+\tr_k(f(\overline{T})\overline{C}H)\right)\delta(x) \nonumber \\
\end{eqnarray}          
As in case of non-diagonal terms of fermions in previous section, we can show from transformation law for tachyon on original system
of $N$ non-BPS branes, that this remaining tachyon field transforms under $U(N-k)\times U(k)$ as :
\[ T \rightarrow gTh^1 , g \in U(N-k), h\in U(k) \]
Then can be easily prove, that terms in (\ref{fermi2}) are invariant under $U(N-k)\times U(k)$ gauge transformations.(under this
transformation, we have: $ E\rightarrow gEg^{-1}, hHh^{-1}  , C\rightarrow hCg^{-1}, B\rightarrow gBh^{-1} f(T) \rightarrow
gf(T)h^{-1}, g\in U(N-k), h\in U(k) $, so that invariance is evident.)

 From string theory point of view, this is interaction term, that we would expect from interaction between string starting and ending on brane and string starting on  brane and ending on antibrane. First term in (\ref{fermi2})reproduces  this interaction. $T^{i\overline{j}}$ corresponds to string starting 
on i-th brane and ending on j-th antibrane, $C^{\overline{j}k}$ corresponds to the string starting on antibrane and ending on brane and $E^{ki}$ comes from string starting and ending on brane. Then we make trace over this factor and we obtain precisely
interaction as we can expect from string theory. We than can say, that first term correspond to the interaction on brane and the
second to the interaction on antibrane. When we write all terms (\ref{Dx}, \ref{Dy}, \ref{pot2}, \ref{fermi2}) together, we obtain function $F$, that contain delta function $\delta(x)$. Now when we insert this term into action for $N$ 9-branes in (\ref{action}), we obtain result:\newpage
\begin{eqnarray}\label{action2}
S=-C\int_{R^{1,8}}(1+\frac{1}{4})\left\{\tr_{N-k}\left( F_{\mu\nu}F^{\mu\nu}+2D_{\mu}XD^{\mu}X+
2i\overline{\theta}(\Gamma^{\mu}D_{\mu}\theta+\right.\right. \nonumber \\
\left.\left.+\Gamma^9[X,\theta]\right)
+4i\tr_{k}(\overline{B}\Gamma^{\mu}\tilde{D}_{\mu}B+\overline{B}\Gamma^9(XB-BX')) \right. + \nonumber \\
\left.+\tr_{k}\left( F'_{\mu\nu}F'^{\mu\nu}+2D_{\mu}X'D'^{\mu}X'
+2i\overline{\theta}'(\Gamma^{\mu}D'_{\mu}\theta'+\right.\right. 
\left.\left.+\Gamma^9[X',\theta'])\right)\right\}\times \nonumber \\
\times \left\{NT^2_v+2\tr_{N-k}(XT-TX')(X'\overline{T}-\overline{T}X) + \right. \nonumber \\
\left.+2\tr_{N-k}(\tilde{D}T^{\mu}\overline{\tilde{D}_{\mu}T}+(-m^2\tr_{N-k}(T\overline{T})+\lambda \tr(T\overline{T})^2)\right. 
\nonumber \\
\left. +\tr_{N-k}(f(T)\overline{B}\theta)+\tr_k(f(\overline{T})\overline{C}\theta')\right\}+I_{WZ} \nonumber\\
\end{eqnarray}
where we have made exchange $ E\rightarrow \theta, \ H\rightarrow \theta'$.

We propose, that this action is correct action for system of $N$ D8-branes and $N-k$ anti D8-branes. Of course, we also expect, that this action is correct action for lower dimensional system of branes and antibranes. We can see, that this action leads to
the action for distinguishable branes and antibranes as was presented in literature in case, when we expand bracket $(1+\frac{1}{4})$ and neglect all terms containing products of terms from first bracket with terms from the second bracket, which can be schematically written as:
\[ S=-\int(1+\frac{1}{4})(F^2+...)(\tilde{D}T^2+...)\sim -\frac{1}{4}\int(F^2+...)-\int(\tilde{D}T^2...)\]

Action (\ref{action2}) reduces in case $k=0$ to the action describing $N$ D8-branes, because than tachyon field $T$ is zero and
the second bracket in (\ref{action2}) reduces to the term $NT^2_v$, from this fact we can obtain value of $T_v$. We know, that in 
unit $2\pi l_s^2=1$, tension of non-BPS brane is $C=\sqrt{2}{g}$
\footnote{We do not include into tension of brane various powers of $\pi$.}
 and from requirement that resulting action describes $N$ D8 branes, so that resulting tension is $T=N/g$, so we obtain condition for vacuum value of tachyon $ T_v$:
\begin{equation}
\frac{\sqrt{2}NT^2_v}{g}=\frac{N}{g} \Rightarrow T^2_v=\frac{1}{\sqrt{2}}
\end{equation}

We can take special case, where action (\ref{action2}) describes one brane and one antibrane. Then the gauge group on 
brane+antibrane system is  $U(1)\times U(1)$.  Firstly, we can ask, what happen for case $T=T_0$, where $T_0$ is now 
minimum of tachyon potential appearing in (\ref{action2}).The vacuum value is $|T_0|^2=T^2_v=\frac{m^2}{2\lambda}$. For constant
tachyon, all derivates are zero and we can deduce  from second bracket of (\ref{action2}) function, from which we obtain
equation of motion for tachyon:
\begin{equation}
F=2T_v^2+2((XT-TX')^2+(AT-TA')^2+V(T))+f(T)(fermions...)
\end{equation}
where the last term contains fermions, but we do not need its precise form.
Then  equation of motion for constant $T$ reduce to :
\begin{equation}
\frac{dF}{dT}=0 \Rightarrow  \overline{T}((X-X')^2+(A-A')^2)+\frac{dV}{dT}+\frac{df(T)}{dT}(....)=0
\end{equation}
We would like to evaluate this equation in point, where $T=T_0$. Then derivation of potential is zero  for this value of tachyon
 from the fact, that vacuum value of $T$ is value minimalising  potential. Because $T$ is nonzero, we obtain solution
of equation of motion in form $ X=X', A=A' $. First condition has consequence, that brane and antibrane in $T_0$ coincide, which 
we can expect from the fact, that between brane and antibrane is attractive force. The second condition means, that gauge
fields must be the same, which again means, that there is no charge for lower dimensional brane. The last thing is case of fermions. Because $f(T)$ is odd function, its derivation is even, we can expect, that is nonzero for $T=T_0$. Then in order to obey
equation of motion, the fermionic term must be zero, so that all spinor fields must be zero in point, where $T=T_0$
\footnote{This conclusion is not precisely correct and right form will be given in the next example.}. We arrive to
the conclusion, that in this point supersymmetric vacuum is recovered. We would like also, that action in this point is zero. 
We see, that all terms in $F$ are zero except $ 2T_v^2+2V(T) $. Vanishing of this term leads to the second condition for vacuum
value of tachyon, that has a form:
\begin{equation}\label{con3}
V(T_0)=T_v^2 \Rightarrow m^4=2\sqrt{2}\lambda
\end{equation}

In the following examples we will show, that this action  correctly reproduces
some properties of  system of branes and antibranes, that are expected on general grounds, and confirm the conclusion, that this action is correct action 
for system of branes and antibranes.

\subsection{Kink solution on pair D8-brane D8-antibrane}

Now we will construct kink solution for tachyon on pair D8-brane D8-antibrane. We use action given in (\ref{action2}) for 
special value of $N=2,k=1$. Then expression in second bracket in (\ref{action2}) has a form(we will be interested mainly
in the second bracket, because only this part contains tachyon.):
\begin{eqnarray}
(2T^2_v+2(XT-TX')^2+2(DT\overline{DT}+V(T))+ \nonumber \\
+f(T)\overline{B}\theta+f(\overline{T})\overline{C}\theta') \nonumber \\
\end{eqnarray}
As was argued in previous part, we expect, that this term is zero in point, where $T=T_0$ is a vacuum value of tachyon. We return
more carefully to the term containing fermions. Because $f(T_0) $ is nonzero, we obtain some condition for them. First possibility is 
$C=B=0$ ,but this it too restrictive and do not leads to the interesting interaction, as we will see. We expand $B=C^*$ in the form: $C=X+iY , B=C^*$,
because they are complex(This property  is reflection of oriented  string.), while
 $\theta,\theta'$ are real (This is due to the fact, that $U(1)$ has not adjoin
representation.). Then we get:
\begin{equation}
f(T_0)C\Gamma^0\theta +f(T_0)B\Gamma^0\theta'=0 \Rightarrow
f(T_0)X\Gamma^0(\theta+\theta')+if(T_0)Y(\theta-\theta')=0
\end{equation}
this equation can be solved with choosing
\begin{equation}
\theta=\theta', X=0
\end{equation}
together with conditions $X=X', A=A'$.

We take tachyon field in the form:
\begin{equation}
T(x,y)=T(x)+\frac{iT(y)\delta(x)}{\sqrt{2}}
\end{equation}
where $x=x^8$ and $y$ are remaining co-ordinates. The factor $1/\sqrt{2}$ has
been taken in order  to obtain correct form of kinetic term for tachyon. We 
choose $T(x)$ in the form of kink solution as in (\ref{T}),
 that is also zero
in point $x=0$, and $T(y)$ is real  function of $y$. The analysis in point outside is similar as before and leads to the previous conditions on massless fields. Due to the condition $A=A'$, that must hold everywhere, the covariant derivate in point $x=0$ leads to the ordinary derivate. Then we obtain as in previous case (in the following, we take $x=x^8$ and $\mu=0,...7$ and again terms with derivate of delta function give zero contributions) :
\begin{equation}\label{k1}
DT\overline{DT}=\delta(x)T^2_v+\frac{1}{2}\partial_{\mu}T(y)\partial^{\mu}T(y)\delta(x)
\end{equation}
With using the condition (\ref{con3}), we see, that constant term cancels with potential except the point $x=0$, where   its contribution is very small with respect to the delta function. In point $x=0$, $T(x)=0$, so that potential has a form:
\begin{equation}\label{k2}
V(T)=\frac{1}{2}(-m^2T(y)^2+\frac{\lambda}{2}T(y)^4)\delta(y)
\end{equation}
The last thing is term with fermionic interaction. At point $x=0$ this  can be written as:
\begin{equation}
(if(T(y))iY\Gamma^0\theta-if(T(y)(-iY\Gamma^0{\theta})\delta(x)=2f(T(y))Y\Gamma^0\theta
\end{equation}
where we have used the fact, that $f$ is real function and where we have included factor $-\frac{1}{\sqrt{2}}$ into definition of $f$ function.
 When we remember, that $Y$ comes from right handed sector in original configuration and $\theta$ comes from left handed sector, than we can rename these fermions as $ Y\rightarrow \theta_R , \theta \rightarrow \theta_L $,  then we obtain final
expression:
\begin{equation}\label{fermi3}
2f(T(y))(\overline{\theta_R}\theta_L)\delta(x)
\end{equation}

Gluing (\ref{k1}, \ref{k2}, \ref{fermi3}) together, we obtain final result from bracket containing tachyon field:
\begin{equation}\label{kfinal}
2\left\{\frac{1}{2}\partial_{\mu}T\partial^{\mu}T+T^2_v+\frac{1}{2}(-m^2T^2+\frac{\lambda}{2}T^4)
+f(T)\overline{\theta_R}\theta_L\right \}\delta(x)
\end{equation}
We can see immediately intriguing similarity with term (\ref{Fa}, \ref{F}). Which can suggests, that this construction leads to lower
dimensional non-BPS D-brane. This will be seen also from analysis the first bracket in (\ref{action2}).

We know, that field theory on brane and antibrane has gauge symmetry $U(1)\times U(1)$, so that field strength is abelian and scalar fields and fermionic fields do not couple to the gauge fields, because $U(1)$ group has not adjoin representation, which also leads to the absence all interaction terms between 
massless fields. First bracket also contain term $\overline{B}\Gamma^{M}\tilde{D}_MB, (M=0,...8)$, that reduce due to the condition $A=A'$ into
$ \overline{B}\Gamma^{\mu}\partial_{\mu}B$, where we have used  well know fact, that all massless fields are independent on $x$.
 The next steps are straightforward. Again, $A_8$ is interpreted as scalar field 
describing fluctuation in $x^8$ direction and $F_{MN}F^{MN}=F_{\mu\nu}F^{\mu\nu}+2\partial^{\mu}X^8\partial_{\mu}X^8$.
From the fact, that  the second bracket in (\ref{kfinal}) contains factor $2$, so that terms for $A,X,\theta_L,\theta_R $ are doubled, we
make redefinition:
\begin{equation}
2A \rightarrow A, 2X \rightarrow X, 2\theta_L \rightarrow \theta_L, \sqrt{2}\theta_R \rightarrow \theta_R
\end{equation}
so that final result is
\begin{eqnarray}\label{final}
S=-C\int_{R^{1,7}}(1+\frac{1}{4})\left( F_{\mu\nu}F^{\mu\nu}+2\partial^{\mu}X^i\partial_{\mu}X^j\delta_{ij}+\right. \nonumber \\
\left. +2i\overline{\theta_R}\Gamma^{\mu}\partial_{\mu}\theta_R+
2i\overline{\theta_L}\Gamma^{\mu}\partial_{\mu}\theta_L \right) \times \nonumber \\
\times \left(\frac{1}{2}\partial_{\mu}T\partial^{\mu}T+T^2_v+\frac{1}{2}(-m^2T^2+\frac{\lambda}{2}T^4)
+f(T)\overline{\theta_R}\theta_L\right) \nonumber \\
\end{eqnarray}
which is a correct action for non-BPS D7-brane as proposed in \cite{Sen}. As a check, consider tachyon condensation on
world-volume of this brane. For simplicity, we take $T$ to be equal to its vacuum value everywhere. We must determine vacuum value of tachyon. From definition, this value minimalise potential, so we have:
\begin{equation}
\frac{dV}{dT}\left|_{T_0}\right.=0 \Rightarrow T_0=\frac{m^2}{\lambda} , V(T_0)=-\frac{m^4}{2\lambda}
\end{equation}
We see, that this value exactly cancels with constant term $T^2_v$,  using (\ref{con3}):
\begin{equation}
T^2_v+\frac{1}{2}V(T_0)=\frac{1}{\sqrt{2}}-\frac{1}{\sqrt{2}}=0
\end{equation}
so that including term $T_v^2/2$ into potential leads to the potential with 
 $V(T_0)=0$, which is correct value of tachyonic potential  for non-BPS D-brane. As a further check, we can proceed in the same way as in section devoted to the construction 
of single brane from unstable brane. We do not mean to repeat the same calculation, but we note only one thing. For tachyon in kink solution, the $F$ function reduces to the form $F=\frac{T^2_o\delta(x)}{2}$, where now we have from potential term
:
$V(T)=-m^2T^2+\frac{\lambda}{2}T^4$ the minimum of the potential: $T^2_0=\frac{m^2}{\lambda}=\frac{2}{\sqrt{2}}$. When we insert
this into (\ref{final}), we obtain action for BPS D-brane with correct tension from $CT^2_0/2=\sqrt{2}\frac{1}{g\sqrt{2}}=\frac{1}{g}$.

We can also consider modding of theory with operator $(-1)^{F_L}$, as in ref.\cite{Sen3}. It was argued, that non-BPS brane is 
invariant under this projection, so that we can look for spectrum of low energy states in string theory living on brane. We start with system of brane+antibrane. As is well know, strings starting and ending on different branes have different CP matrix. We can 
take $CP$ matrices in the form:
\begin{eqnarray}
p-p=\left(\begin{array}{cc} 1 & 0 \\
                                  0 & 0 \\ \end{array} \right) , \ p-\overline{p}=\left(\begin{array}{cc} 0 & 1 \\
                                                                                                                                      0 & 0 \\ \end{array} \right) \nonumber \\
\overline{p}-p=\left(\begin{array}{cc} 0 & 0 \\
                                                 1 & 0 \\ \end{array}\right), \ \overline{p}-\overline{p}=
\left(\begin{array}{cc} 0 & 0 \\
                                 0 & 1 \\ \end{array}\right)  \nonumber \\
\end{eqnarray}

States, that are invariant in system brane+antibrane, are with $CP$ matrices $I$ and $\sigma_1$, whereas states with $\sigma_3$ and $i\sigma_2$ are projected out
\footnote{$\sigma_1=\left(\begin{array}{cc} 0 & 1\\
                                                               1 & 0 \\ \end{array}\right),
i\sigma_2=\left(\begin{array}{cc} 0 & 1 \\
                                                -1 & 0 \\ \end{array}\right)$.}
 We take system brane+antibrane in IIA theory. Then combination
$T-\overline{T}$ is projected out, which is equivalent to the condition: \[T-\overline{T}=0\]
When we write $T=A+iB$, we obtain $B$=0, so that tachyon is real after projection.
The same argument leads to the condition for spinor $B$ coming from $p-\overline{o}$ sector, results into:
 \[B-\overline{B}=0 \] leading to the result, that this spinor must be real.  

From the fact, that states with $\sigma_3$ are odd, we have condition: 
\[A-A'=0, X-X'=0, \theta-\theta'=0\]
, so that these fields must equal. As a result,   we obtain in action
for system brane+antibrane doubled number of terms for massless fields, 
therefore we must 
redefine these fields:\[\sqrt{2}A \rightarrow A, \sqrt{2}X \rightarrow X, \sqrt{2}\theta \rightarrow \theta,  T \rightarrow \frac{T}{\sqrt{2}}\]
(we have scaling in tachyon and other fields to obtain kinetic terms with standard normalisation)as well as $ B \rightarrow \theta'_L$, because modding with $(-1)^{F_L}$ leads to the transformation of  IIA to IIB.
As a result, we obtain action for non-BPS D-brane in IIB theory:
\begin{eqnarray}\label{final2}
S_{IIB}=-C\int_{R^{1,7}}(1+\frac{1}{4})\left( F_{\mu\nu}F^{\mu\nu}+2\partial^{\mu}X^i\partial_{\mu}X^j\delta_{ij}+\right. \nonumber \\
\left. +2i\overline{\theta_L'}\Gamma^{\mu}\partial_{\mu}\theta_L'+
2i\overline{\theta_L}\Gamma^{\mu}\partial_{\mu}\theta_L \right) \times \nonumber \\
\times \left(\frac{1}{2}\partial_{\mu}T\partial^{\mu}T+T^2_v+\frac{1}{2}(-m^2T^2+\frac{\lambda}{2}T^4)
+f(T)\overline{\theta'_L}\theta_L\right) \nonumber \\
\end{eqnarray}
We can again modding out this theory with $(-1)^{F_L}$, that leads in the bulk to the transformation of IIB into IIA theory. Because 
non-BPS D-brane does not carry RR charge, it is invariant under this transformation. As was explained elsewhere \cite{Sen3}, that
state with $CP$ factor $I$ are even under this projection and state with $\sigma_1$ are odd. Then we must take tachyon in action
to be zero: $T=0$, as well as: $\theta_L'=0$. As a result, the last bracket in (\ref{final2}) is equal to $\frac{1}{\sqrt{2}}$ and we obtain correct tension for BPS brane:
\[\frac{C}{\sqrt{2}}=\frac{1}{g} \]
and the first term leads to the ordinary action for BPS D-brane.
 
Now we must return to the  interaction of type $g(T)\theta_L\theta_L$. In the firs section, we have supposed, that this term is not important in our calculation. This is possible
only in case, when this function obey following conditions. Firstly, in order to obtain from non-BPS brane BPS brane by operation $(-1)^{F_L}$, we must pose the 
condition  $g(T=0)=0$, in order to have $F=\frac{1}{\sqrt{2}}$. Secondly, we should also
have condition $g(T_0)=0,\frac{dg}{dt}\left|_{T=T_0}=0\right.$.When the letter condition was not satisfied, we would have to
impose condition $\theta_L=\theta_R=0$ and we would not have the correct actions for BPS branes. Because we have seen, 
that from our analysis we are able to obtain actions for all branes in Type IIA theory, we can suppose, that this function obeys these
conditions. It is not difficult to find such a function,  for example $g(T)=T^2(T^2-T^2_0)^2$. In the following, we will suppose,
that $g$ obeys previous conditions and we will not implicitly write this term into action.

 We have seen, that from tachyon condensation in the form of kink solution on 
world-volume of non-BPS D9 branes, we have obtained action for D8-branes and D8-antibranes. We have also seen, that further tachyon condensation on the pair D8-brane +D8-antibrane leads to non-BPS D7-brane, as can be expected on general grounds
\cite{witen,Horava}. And finally, the last tachyon condensation leads to BPS
D6-brane. In this ``step by step'' construction we would be able to construct
the whole spectrum BPS and non-BPS D-branes in IIA theory, again with completely agreement with \cite{Horava}. But certainly it would be nice to have a 
contact with one step construction, proposed in \cite{Horava}. In the last 
section, we show first step in this direction. We will construct tachyon
vortex solution on pair D8-brane+D8-antibrane.
\subsection{Vortex solution on pair brane+antibrane}

Now we proceed to the question of construction of vortex solution on system 
D8-brane and D8-antibrane. The gauge theory describing this system is 
$U(1)\times U(1)$ and action has a form:
\begin{eqnarray}
S=-C\int_{R^{1,7}}(1+\frac{1}{4})\left(F_{MN}F^{MN}+F'_{MN}F'^{MN}+
\right. \nonumber \\
\left.+ 2\partial_MX\partial^MX+2\partial_MX'\partial^MX'+
2i\overline{B}\Gamma^M\partial_MB+2i\overline{\theta}\Gamma^M\partial_M\theta+
2i\overline{\theta}'\Gamma^M\partial_M\theta'\right)\times \nonumber \\ 
\times \left(2T^2_v+2(D_MT\overline{DT}^M+V(T)     
+(XT-TX)^2) 
+f(T)\overline{B}\theta+
f(\overline{T})B\Gamma^0\theta'\right) \nonumber \\ 
\end{eqnarray}

Now we would like to construct vortex solution, where tachyon field forms a vortex in space $R^2$ on world-volume of brane+antibrane pair (we divide world-volume of this pair as $R^{1,7}=R^2\times R^{1,5}$ and we take cylindrical co-ordinates in the $R^2$ space $r,\phi$), that has a form:
\begin{equation}
T=T_ve^{i\phi}
\end{equation}
Strictly speaking, the tachyon should approach to this solution in spatial infinity, but we would like solution, that is in vacuum value in the whole plane 
$R^2$ except some small region around point $r=0$. Equation of motion for tachyon is:
\begin{equation}
\frac{\delta F}{\delta\overline{T}}=-D_MD^MT+\frac{dV}{d\overline{T}}
+\frac{1}{2}\frac{df}{d\overline{T}}B\Gamma^0\theta'+T(X-X')^2=0
\end{equation}
and the same for $\overline{T}$. We will make the same analysis as in previous cases of kink solution, but there is an important difference. In kink solution, we have specified only real part of tachyon, so that imaginary part remained 
free, so we have generally obtained non stable configurations after tachyon condensation. In vortex solution, the tachyon is fixed completely, even in point $r=0$,
so that there is no place for remaining tachyon modes, so that we should expect, that we obtain stable configuration  following previous results 
\cite{Sen,Horava,witen} we can expect, that resulting configuration is 
supersymmetric D6-brane.

As in usually, we start with points in plane $R^2$, where tachyon is in vacuum value. Again, because tachyon vacuum value is nonzero, we obtain condition for fields describing fluctuation in 9-th direction: 
\begin{equation}\label{conVa}
X-X'=0
\end{equation}

From the requirement of continuity of massless field, this constraint must be
obeyed on the whole plane, so that brane and antibrane system has only 
degrees of freedom corresponding to the transverse fluctuation of the system as it would be the single object. 

Now we return to the equation $\frac{df}{2d\overline{T}}B\Gamma^0\theta'$ (and 
the analogous equation for $\overline{T}$) in the point, where $T$ is in vacuum
value. Because these derivation is nonzero, we obtain conditions for spinor fields (the situation is the same as in case of kink solution, so we do not mean to go into details). Due to the fact, that $f$ is real function, we can write this derivation as follows:
\[\frac{df}{d\overline{T}}=g-ih, \frac{df}{dT}=g+ih\]
and in the same way: $B=X+iY , B^*=X-iY$.
Then we obtain :
\begin{equation}
gX\Gamma^0(\theta+\theta')+ih\Gamma^0(\theta-\theta')+igY\Gamma^0(\theta'
-\theta)+hY\Gamma^0(\theta+\theta')=0
\end{equation}
It is clear, that solution of this equation is
\begin{equation}\label{conVf}
X=0, Y=0 , \theta=\theta'
\end{equation}
implying, that spinor arising from string sector connecting brane and antibrane disappears and two spinors from string sector living only on brane and antibrane combine to one independent spinor. Again, this implies existence of supersymmetric brane.

The equation $\frac{dV}{dT}$ is zero in point $|T|=T_0$ and in point $x=0$, 
where we expect tachyon field to be zero, is zero as well, because value of
tachyon $T=0$ is also minimum of potential.But in the small region around point $x=0$ the behaviour of this potential will be nontrivial, because tachyon goes
from $T=T_0$ to the $T=0$, this function must change very rapidly, so we expect, that infinity from this potential cancel the infinity from derivation of tachyon, so that equations of motion are obeyed.

The last term is $D_MD^MT=0$, that leads to the condition $D_MT=0$. For
$\mu=0,...5$, we obtain (due to the fact, that tachyon is not function of 
these co-ordinates):
\begin{equation}\label{conV}
(A-A')_{\mu}T=0 \Rightarrow A_{\mu}=A'_{\mu}
\end{equation}
again we have only one independent gauge fields in this direction. 
For the directions in plane $R^2$, the situation is more subtle. Because
tachyon has nonzero twist, the requirement $D_{\phi}T=0$ leads to the
condition \[A_{\phi}-A'_{\phi}=\frac{1}{r}\] and condition $D_rT=0$ in point
outside $r=0$ leads to the condition $A_r-A_r'=0$. Because tachyon is fixed also in point $x=0$, this functions are fixed in this points as well, so we have
conditions: $A_r-A'_r=0$ holding everywhere and $A_{\phi}-A'_{\phi}$ is fixed in some nontrivial way. In other word, even if the field $A_{\phi}-A'_{\phi}$ and its derivation
is nonzero in point $x=0$, this field are not free, so that they appear in resulting  action as a constant, that can be taken out, but the nontrivial behaviour of $A_{\phi}-A'_{\phi}$ is important in WZ term, as we will see.

On the other hand, fields $A^S_r=A_r+A'_r , A^S_{\phi}=A_{\phi}+A'_{\phi}$ are
 free, which  leads to the important consequence.
To conclude, in order to obey equations of motion for tachyon, we have obtained
number of constraint on massless fields living on brane and antibrane and also we have obtained the form of the second bracket in the action:
\begin{equation}\label{delta}
 (...)=2T^2_v\delta(r)=2T^2_v\delta(x)\delta(y)
\end{equation}
As was pointed in previous part, the only nontrivial fields coming from gauge fields with indexes $r,\phi$ are fields $A^S_{x,y}$, where we again use Cartesian co-ordinates in plane $R^2$. To explain this fact more carefully, we rewrite kinetic terms for gauge fields as follows (we include factor $2$ from 
(\ref{delta})):
\begin{eqnarray}
2(F_{MN}F^{MN}+F'_{MN}F'^{MN})=2F_{\mu\nu}F^{\mu\nu}+2F'_{\mu\nu}F'^{\mu\nu}+
 \nonumber \\
+4F_{\mu i}F^{\mu i}+4F'_{\mu i}F'^{\mu i}+4F_{xy}F^{xy}
+4F'_{xy}F'^{xy}= \nonumber \\
=F^s_{\mu\nu}F^{s\mu\nu}+F^a_{\mu\nu}F^{a\mu\nu}+2(F^s_{\mu j}F^{s\mu j}
+F^a_{\mu j}F^{a\mu j})+2(F^s_{xy}F^{s xy}+F^a_{xy}F^{a xy}) \nonumber \\ 
\end{eqnarray}
where for the time being $ j=8,7$ and where we have introduced combinations:
$A^s_M=A_M+A'_M , A^a_M=A_M-A'_M $. 
This expression simplifies dramaticly, when we use previous results. Firstly,
factor $ F_{\mu\nu}^a$ is zero due to the condition (\ref{conV}). Secondly, 
term $ F^a_{\mu j}$ is zero as well, because contains fields, that are 
completely fixed with tachyon condensation and this is independent 
on $\mu=0,..6$
co-ordinates. And finally last bracket is zero as well, because symmetric 
combination of field is independent on $x,y$ co-ordinates as other massless fields and antisymmetric combination is completely fixed with tachyon,
so that has not any dynamical effect and consequently can be put away from
the Lagrangian. As a result, we obtain following kinetic term (We rename 
$A^s_{\mu}=A_{\mu}, A^s_{x,y}=X_{x,y}$ and we include kinetic term for scalar
fluctuation in 9-th direction, which appears only in symmetric combination
due to the (\ref{conVa})):
\begin{equation}
F_{\mu\nu}F^{\mu\nu}+2\partial_{\mu}X^i\partial^{\mu}X^j \delta_{ij}
\end{equation}
where $i,j=7,8,9$. We can also introduce symmetric and antisymmetric combinations for fermions and with using (\ref{conVf}) we find out, that only symmetric combination survives.
Finally, with using the fact, that $CT^2_v=\frac{1}{g}$, we obtain action for supersymmetric D6-brane:
\begin{equation}
S=-\frac{1}{g}\int_{D6}(1+\frac{1}{4})\left(F_{\mu\nu}F^{\mu\nu}+
2i\overline{\theta}\Gamma^{\mu}\partial_{\mu}\theta+
2\delta_{i,j}\partial_{\mu}X^i\partial^{\mu}X^j \right)
\end{equation}

As a next thing we must return to the WZ term. We will follow ref.
\cite{Wilkinson}, where WZ term for system brane+antibrane was presented. As
was explained in this paper, in case of tachyon condensation into vortex
solution, the tachyon field do not contribute the RR charge and we obtain
 following coupling:
\begin{equation}
I_{WZ}=k\int_{R^{1,7}}C_{8}\wedge(e^F-e^{F'})
\end{equation}
where $k$ is unknown constant.
When we expand the exponents, we obtain:
\begin{eqnarray}
I_{WZ}=k\int C_7\wedge (F-F')+\frac{k}{2}\int C_5\wedge (F-F')\wedge(F+F')+ \nonumber \\
+\frac{k}{6}\int C_3\wedge (F-F')\wedge (F\wedge F+F\wedge F'+F'\wedge F')+\nonumber \\
+\frac{k}{24}\int C_1\wedge(F-F')\wedge (F+F')\wedge (F\wedge F+F'\wedge F') \nonumber \\
\end{eqnarray}
It is well known, that integral$ \int(F-F')$ has value $2\pi$ for  tachyon vortex, so that first term express coupling D6-brane to the $C_7$ form and
consequently  $k$ is proportional to RR charge of six brane (Again, we do not
worry about various factors of $\pi$, which can be easily found with help of  dimensional analysis.).

The other terms are more subtle. The second term gives coupling: $\frac{k}{2}
\int C_5 (F^s)$, which could suggests, that resulting brane carry one half 
of $C_5$ charge. The other terms give the same results. we do not know how this
solve, maybe we should use the  result in \cite{Wilkinson}, where WZ term for
system brane and antibrane is expressed in form of super connection.

To sum up, we have obtained action for D6-brane as a result
of tachyon condensation in form of vortex solution on world-volume of  system D8-brane+D8-antibrane. We have seen, that this solution carries correct value
of charge for coupling with $C_7$, but question of charges for lower RR forms
remains open.

\section{Conclusion}

In previous parts we have seen, that we are able to construct action for all 
D-branes in IIA theory via tachyon condensation in the procedure named 
``step by step'' construction, which was  proposed in \cite{Horava}. Then we
have seen, that we can construct lower dimensional brane in form of vertex
solution on world-volume of system brane+antibrane, which certainly could
be used to construction all D-branes in IIB theory, following \cite{witen}.

As a next step in this analysis, is construction more general vortex solutions 
directly in world-volume of non-BPS D9-branes in IIA theory or in world-volume
of system D9-branes and D9-antibranes in IIB theory. Further it would be
interesting to study more general configurations, where D-branes are in general positions and the configurations with branes and antibranes, that  are not coincided. It would  be also interesting, following \cite{Horava}, to study relation 
between tachyon condensation and Matrix theory. It seams to us, that tachyon 
could  have some connection with the origin of the branes, with breaking of 
supersymmetry, and consequently, through construction of D0-branes and their
using in Matrix theory, with M theory and holography.

As the last thing we must say a few words about WZ terms. We have seen, that for tachyon condensation on N non-BPS D9-branes, we obtain WZ term, corresponding to the difference of WZ terms for D8-branes and D8-antibranes. This is certainly a good result, but seams to us, that there is still something missing, 
because  instability of resulting configuration does not appear in this term. 
As a result, through further tachyon condensation on this system, we obtain 
unstable non-BPS D7-brane and from our analysis was clear, that this brane does
not carry charge of any lower dimensional RR form. Again, in some sense this is correct
result, because non-BPS D-brane does not carry RR charge, but we know, that
further tachyon condensation leads to stable BPS D-brane, but due to the absence of  coupling between tachyon and RR form, we cannot obtain coupling between 
BPS brane and RR form. In other words, we do not see relation 
between WZ term proposed in \cite{Wilkinson} and in \cite{Billo}.
Better understanding of this relation could  lead to the explanation, why we
do not have correct value of charges for lower dimensional branes in D6-brane,
as we have seen in the last section. We mean, that resolution of this
puzzle is very important for further analysis.

\end{document}